\documentclass[10pt,twocolumn, journal]{IEEEtran}

\usepackage{array,graphicx,subfigure,epsfig,amsmath,amsthm,amssymb,amsfonts,multirow,hhline,enumerate,
cite,epstopdf,array,delarray,verbatim,color,caption}
\usepackage{psfrag}
\usepackage{tabularx}
\usepackage{algpseudocode,algorithm,algorithmicx}
\usepackage{blindtext}
\usepackage{graphicx}
\DeclareGraphicsExtensions{.pdf,.png,.jpg}

\begin{document}

\title{Multiple Access for 5G New Radio:\\ Categorization, Evaluation, and Challenges}

\author{Hyunsoo Kim,~\IEEEmembership{Student~Member,~IEEE,}
        Yeon-Geun Lim,~\IEEEmembership{Student~Member,~IEEE,}\\
        Chan-Byoung Chae,~\IEEEmembership{Senior~Member,~IEEE}, and
        Daesik Hong,~\IEEEmembership{Senior~Member,~IEEE}
\thanks{H. S. Kim and D. S. Hong are with School of Electrical and Electronic Engineering, Yonsei University, Korea. Y.-G. Lim and C.-B. Chae are with School of Integrated Technology, Yonsei University, Korea (e-mail: {hyunsookim, yglim, cbchae, daesik}@yonsei.ac.kr). Corresponding author is {D. S. Hong.}}
}

\maketitle

\begin{abstract}

Next generation wireless networks require massive uplink connections as well as high spectral efficiency. It is well known that, theoretically, it is not possible to achieve the sum capacity of multi-user communications with orthogonal multiple access. To meet the challenging requirements of next generation networks, researchers have explored non-orthogonal and overloaded transmission technologies--known as new radio multiple access (NR-MA) schemes--for fifth generation (5G) networks. In this article, we discuss the key features of the promising NR-MA schemes for the massive uplink connections. The candidate schemes of NR-MA can be characterized by multiple access signatures (MA-signatures), such as codebook, sequence, and interleaver/scrambler. At the receiver side, advanced multi-user detection (MUD) schemes are employed to extract each user's data from non-orthogonally superposed data according to MA-signatures. Through link-level simulations, we compare the performances of NR-MA candidates under the same conditions. We further evaluate the sum rate performances of the NR-MA schemes using a 3-dimensional (3D) ray tracing tool based system-level simulator by reflecting realistic environments. Lastly, we discuss the tips for system operations as well as call attention to the remaining technical challenges.

\end{abstract}


\section{Introduction}
\IEEEPARstart{W}{ireless} communication systems have evolved through one generation supplanting its predecessor. Accompanying this evolution has been the progression of multiple access (MA) technologies. For example, from the first generation (1G) to the 4G-LTE,
the conventional communication systems orthogonally assigned radio resources to multi-users in time, frequency, and code domains. Experts believe that by 2020 (5G), mobile data traffic will have grown a thousand-fold (1000x)~\cite{5GMA}. Explosive growth in data traffic will be propelled by the Internet of things (IoT) and massive machine-type-communications (mMTC). At the ITU-R WP 5D meeting, the target number of the maximum link connections was determined to be one million per square kilometer~\cite{ITU}. The orthogonal multiple access (OMA) system, however, has limited capability in supporting the massive devices. Regarding sum capacity, it is well known that the OMA system cannot approach the Shannon limit~\cite{5GMA}. A great number of industrial and academic researchers have shifted the focus of their studies from \emph{orthogonality} to \emph{non-orthogonality}. In fact, in the Third Generation Partnership Project (3GPP), NTT DOCOMO introduced the power domain non-orthogonal multiple access (NOMA). The NOMA allows multiple users to share the same radio resources~\cite{NOMA}.

Another movement has proposed various non-orthogonal and overloaded transmission techniques known as new radio multiple access (NR-MA) schemes. The candidates of NR-MA can be characterized by multiple access signatures (MA-signatures), which are identifiers to distinguish user-specific patterns of the data transmissions. The candidates of NR-MA can be grouped into three categories, i.e., i) codebook-based schemes~\cite{SCMA, SCMA_Pattern, PDMA}, ii) sequence-based schemes~\cite{MUSA, MUSA_Rx, NCMA, NOCA, GOCA_RDMA}, and iii) interleaver/scrambler-based schemes~\cite{GOCA_RDMA, IDMA, IGMA, RSMA}. Here, advanced multi-user detection (MUD) algorithms are considered so as to recover superposed signals of NR-MA. In previous studies, however, performance evaluation with different system parameters have been carried out scheme by scheme, making it difficult to compare performances fairly or to extract meaningful insights into system design.

In this article, we elaborate on the basic principles of the NR-MA schemes. Then, through the categorization, we precisely explain the key features and differentiated points among the schemes. In an effort to evaluate how well the candidate schemes perform at overloading users and at sustaining inter-user interference, we conduct link-level simulations under common conditions such as transmission power, the number of resource blocks and superposed users' signals, and channel environments. Furthermore, through 3-dimensional (3D) ray tracing-based system-level simulations, we evaluate the sum rate performance of the NR-MA schemes in realistic environments. To the best of our knowledge, this is the first work that fairly evaluates potential NR-MA schemes being discussed for 5G. Finally, we discuss meaningful insights into system design and point out research challenges to operating NR-MA schemes in practice.

\begin{table*}[t]
\caption{Categories of new radio multiple access schemes.}
\centering
\includegraphics[width=\textwidth]{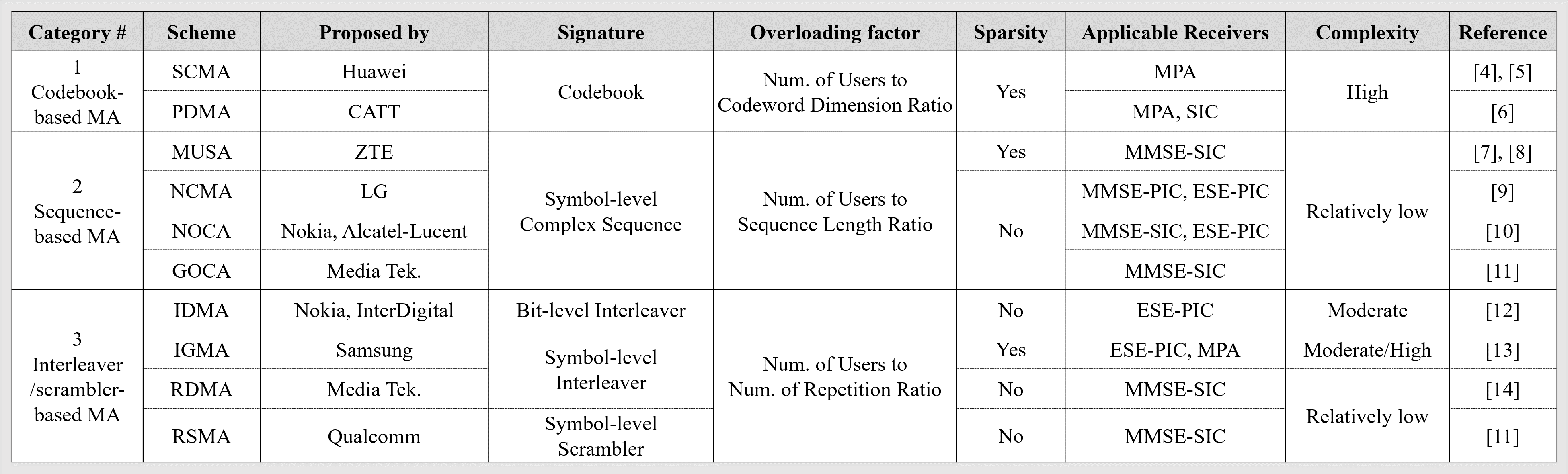} \vspace{-2mm}
\label{t_1}
\end{table*}

\begin{figure*}[t]
\centering
\includegraphics[width=\textwidth]{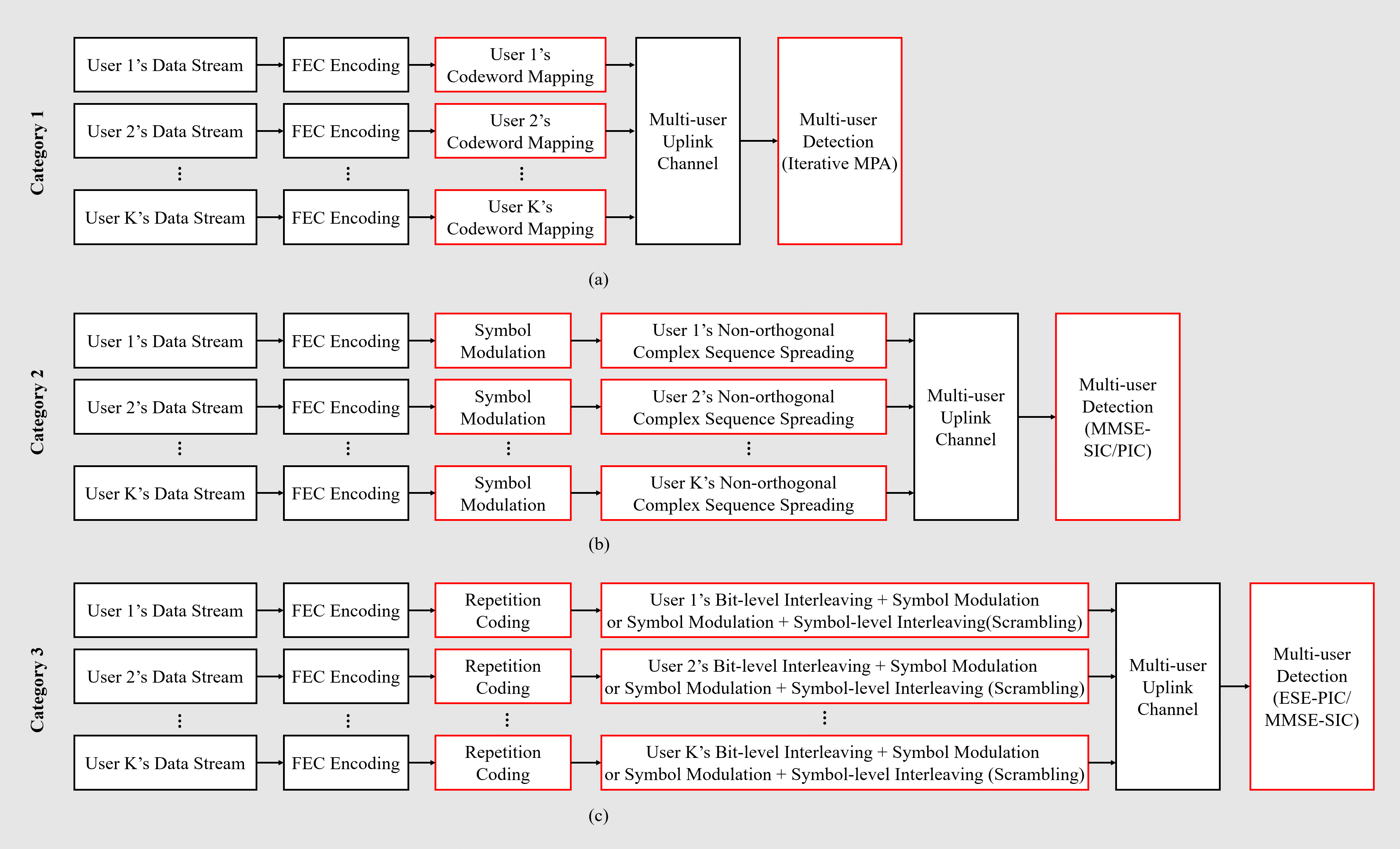}
\caption*{\textbf{Figure 1.} Common system block diagrams for a) codebook-based MA, b) sequence-based MA, and c) interleaver/scrambler-based MA.} \vspace{3mm}
\label{fig_1}
\end{figure*}

\section{Overview of New Radio Multiple Access}

\subsection{Basic Principles}
NR-MA schemes may be described as a user overloading technology that expands a capacity region. The key is how well a large number of users' signals are superimposed and recovered within a controllable and acceptable amount of interference. An important metric for the NR-MA in this regard is the overloading factor defined as the ratio of the number of overloaded signals to that of orthogonal resource grids.

The candidate schemes proposed in the 3GPP share two common features. One, a transmitter spreads data by one or more MA-signatures. Two, a receiver employs advanced MUD algorithms with moderate computational complexity. In the former, the signatures include power, codebook, sequence, interleaver, scrambler, etc. The signal spreading can be operated at either the bit-level, the symbol-level, or at both. By user-specific spreading and resource mapping, it makes each user's signal distinguishable and more robust to inter-user interference. The second feature shared by the candidate schemes is that a receiver employs advanced MUD algorithms with a moderate computational complexity. In the latter, a powerful forward-error-correction (FEC) coding makes up for the insufficient decoding performance of the MUD algorithms.

\subsection{Categorization}

In this subsection, we categorize the candidate schemes introduced in 3GPP, and discuss their common architectures and key features. As shown in Table~\ref{t_1}, the technologies could be categorized according to the predominant use of a MA-signature. Explanations about each category are given below.

\newpage

\begin{itemize}
\item {Category 1: Codebook-based MA}
\end{itemize}

As illustrated in Fig.~1a, the key feature of codebook-based MA schemes is the direct mapping of each user's data stream into a multi-dimensional codeword in a codebook. The codeword has two characteristics: i) signal spreading to obtain diversity/shaping gain, and ii) `zero' elements in a codeword to suppress inter-user interference in a sparse manner. The positions of zero elements in different codebooks are distinct so as to avoid a collision of any two users~\cite{5GMA}. In this scheme, the spreading factor is equivalent to the dimension of a codeword, and the overloading factor is determined by the ratio of the number of multiplexed users to the spreading factor. To recover users' data streams effectively, the scheme adopts an iterative message passing algorithm (MPA) as a near-optimal solution~\cite{SCMA}. Since the MPA receiver is based on maximum likelihood (ML) detection, it has high--relative to other categories--computational complexity compared to other categories.

Two schemes that belong to this category are sparse code multiple access (SCMA) and pattern division multiple access (PDMA)~\cite{SCMA, SCMA_Pattern,PDMA}. The main difference between SCMA and PDMA is the resource-utilization pattern. SCMA codewords in all codebooks have the same number of zero/non-zero elements as a regular pattern. In contrast, the PDMA system allocates a different number of non-zero elements by considering each user's channel state. For example, to obtain a high diversity gain for a user with a weak channel gain, the PDMA can assign a codebook with more non-zero elements than those of another codebook. However, the codebook optimization of PDMA is highly complicated, and remains an open problem. Another differentiated point is that PDMA can utilize a power domain MA-signature. Similar to power domain NOMA, the intended near-far effect from power control can help eliminate interference effectively.

\begin{itemize}
\item {Category 2: Sequence-based MA}
\end{itemize}

Novel sequence-based MA techniques utilize non-orthogonal complex number sequences to overlap multi-user signals as shown in Fig.~1b. This method contrasts with that of using orthogonal pseudo noise sequences in a code division multiple access (CDMA) system. Similar to SCMA, the overloading factor is determined by the length of the spreading sequence and the number of the overloaded users. A family of complex sequences with short lengths is chosen to enable a simple multi-user interference cancellation~\cite{MUSA}. For this, a minimum-mean-square-error with successive/parallel interference cancellation (MMSE-SIC/PIC) schemes have been considered for applicable receivers~\cite{MUSA_Rx}. As the MMSE-SIC/PIC are linear-type receivers, they would be advantageous in terms of computational complexity.

The key issue in sequence-based MA is how to design and assign non-orthogonal sequence sets to users. In multi-user shared access (MUSA), the real and imaginary parts of sequence elements are randomly generated from \{-1, 0, 1\}. Thanks to the zero elements in the sequences, inter-user interference is efficiently mitigated as in the codebook-based MA. Non-orthogonal coded multiple access (NCMA) obtains non-orthogonal sequences by solving a Grassmannian line packing problem~\cite{NCMA}. It is proposed non-orthogonal coded access (NOCA) to utilize the LTE sequences defined for uplink reference signals~\cite{NOCA}. Group orthogonal coded access (GOCA) exploits a dual-sequence: a non-orthogonal sequence for group separation and an orthogonal sequence for the user separation within a group~\cite{GOCA_RDMA}.

\begin{itemize}
\item {Category 3: Interleaver/Scrambler-based MA}
\end{itemize}

Figure~1c illustrates a simplified block diagram for an interleaver/scrambler-based MA system. The repetition coding part determines a spreading factor. The interleaver/scrambler part makes diverse superposition patterns, and then obtains the interference averaging effect. The representative technology of interleaver/scrambler-based MA is interleave-division multiple access (IDMA). In an IDMA system, through a user-specific bit-level interleaver, inter-user interference is suppressed by overlapped signal experiences. While the schemes in Categories 1 and 2 spread data in small units using short codewords/sequences, the interleaver enables dispersion of data across a long signal stream. With the spread signals widely distributed, it is difficult to apply an MMSE filtering and an MPA due to the high computational complexity at the receiver side. In this regard, a receiver of IDMA exploits an elementary signal estimator with PIC (ESE-PIC), which  permits chip-by-chip soft interference estimation and cancellation with a moderate computational complexity~\cite{IDMA}.

By expanding the IDMA system, a sparse symbol-mapping precess known as grid mapping is supplemented by interleave-grid multiple access (IGMA). After the bit-level interleaver and symbol modulation, `zero' symbol padding and resource mapping algorithms are additionally operated. In a repetition division multiple access (RDMA), researchers have introduced a user-specific cyclic repetition pattern for the purpose of avoiding highly correlated interleaving patterns among users~\cite{GOCA_RDMA}. Rather than using the interleaving technique, resource spread multiple access (RSMA) relies on a combination of a low-rate channel coding scheme and a user-specific scrambling. Without the joint decoding process at the receiver side, RSMA has the potential to allow grant-free transmission and asynchronous multiple access~\cite{RSMA}.

\section{Performance Evaluation for Uplink Scenario}
For each category, we select the base schemes that well represent the key features: SCMA, MUSA, and IDMA.\footnote{Note that candidate schemes have different characteristics despite being in the same category. In this article, however, we focus on fundamental and base schemes to make useful observations and identify system design insights for each category.} Since, in previous studies, experiments were performed with different system parameters scheme by scheme, the experiments were limited in their capacity to show how the schemes' performances compared with one another or to provide insight into system design. In this article, we attempt to overcome such limitations by evaluating the schemes under the same conditions at both the link- and system-level. The detailed system parameters and assumptions are described in Table~\ref{t_2}. 

\begin{figure*}
\centering
\includegraphics[width=\textwidth]{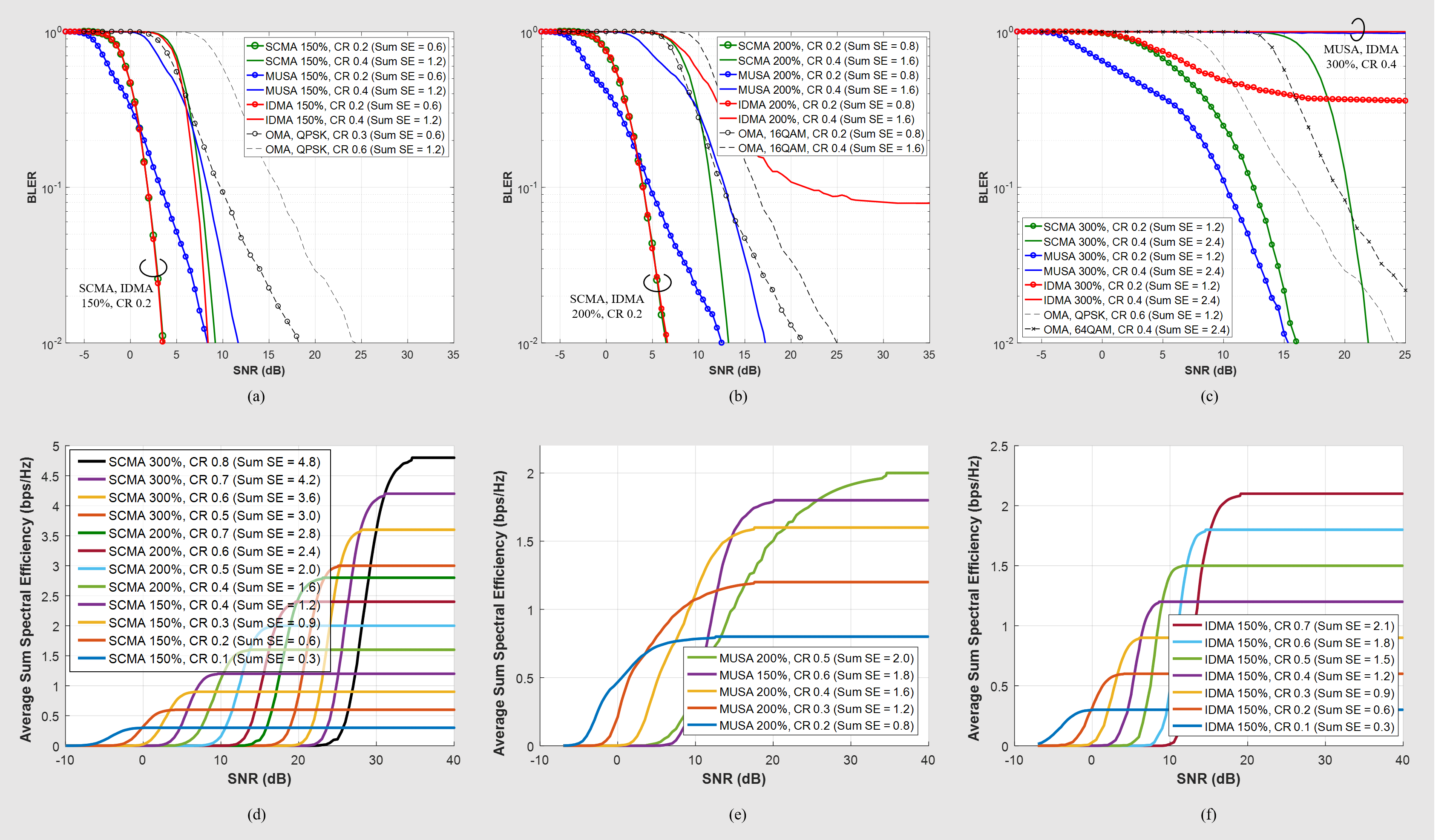}
\caption*{\textbf{Figure 2.} Performance comparisons of candidate MA schemes: a), b), and c) describe BLER curves with overloading factors 150, 200, and 300~\%, respectively. Average sum spectral efficiency (SE) with various overloading factors and code rates (CR): d), e), and f) are results for SCMA, MUSA, and IDMA, respectively.}
\label{fig_2}
\end{figure*}

\begin{table}
\caption{Simulation parameters for uplink scenarios.}
\centering
\includegraphics[width=\columnwidth]{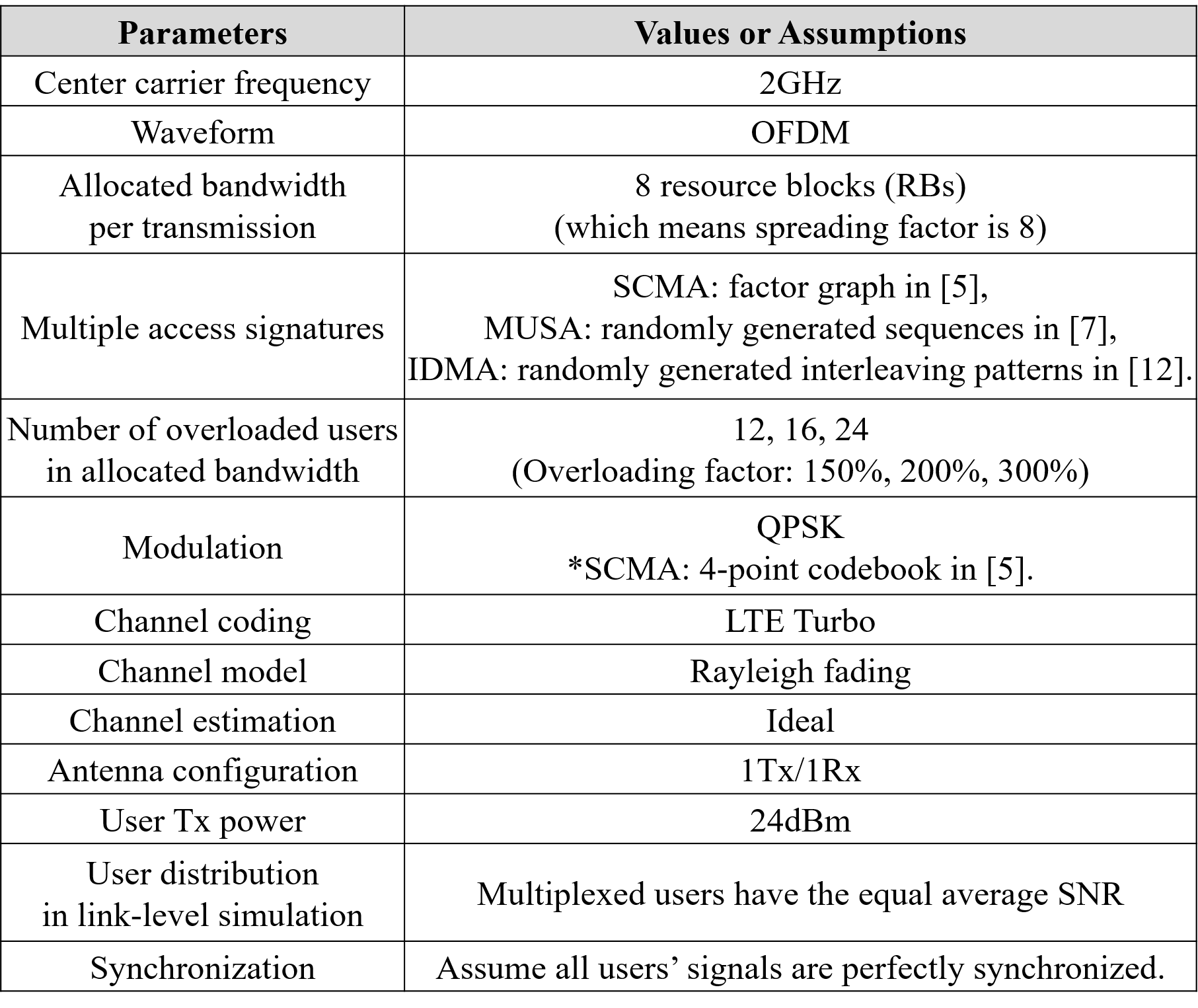}\vspace{-10mm}
\label{t_2}
\end{table}

\subsection{Link-level Simulation} \label{sub}
\begin{itemize}
\item {Comparisons of BLER Performance}
\end{itemize}

Figures~2a-2c show the block error rate (BLER) performance with diverse overloading factors and code rates. The performance of the OMA system is represented as baselines. Without loss of generality, we assume that the received power per resource element and the spectrum efficiency per user is kept the same for all MA schemes including the OMA. For example, when the spreading factor is eight, the power of each chip is normalized by $1/8$. In terms of spectral efficiency, to compare a 200~percent overloaded NR-MA scheme with QPSK, the OMA system sets 16QAM to serve users doubly~fast.

As displayed in Fig.~2a, all candidates outperform OMA when users are overloaded with 150~percent. Especially, the BLER curves of SCMA and IDMA are almost overlapped at the code rates~0.2 and 0.4. Moreover, due to the diversity gain from sparse codeword mapping or bit stream interleaving, the BLER curves have steeper slopes than the OMA curves. MUSA contrarily shows a different tendency. Unlike in SCMA and IDMA, in MUSA the MMSE linear receiver is adopted to mitigate inter-user interference. It reduces the power of the desired signal as well as suppresses the interference; as a result, diversity gain loss occurs.

Figure~2b exhibits the 200~percent overloading case in which IDMA has a unique tendency. The BLER curve of IDMA is saturated when a moderate code rate (0.4) is applied. This is because IDMA treats the inter-user interference as a noise in the initial stage of the ESE-PIC receiver~\cite{IDMA}. If the total interference exceeds the permissible amount in this stage, the superposed signals experience severe performance degradation due to the error propagation in the PIC stage. Even so, IDMA can overcome this problem with the aid of powerful FEC coding with a low rate (0.2) showing a similar performance compared to the SCMA case.

Figure~2c verifies how much interference each technique can endure. SCMA is able to withstand well highly aggregated interference with both the low and moderate code rates. MUSA, on the other hand, does not work with a high overloading factor (300~percent) and a moderate code rate. Since the spreading factor is much smaller than the number of the overloaded users, the effective channel matrix for the MMSE receiving filter does not have rank enough to distinguish a larger number of users' signals~\cite{MUSA_Rx}. This means that it is more difficult to separate each user's signal as the overloading factor increases.\footnote{Note that even though the rank of the effective channel matrix can be extended by using multiple receiving antennas, it leads to an unfair comparison by using more orthogonal space domain. In this article, additional orthogonal spatial resources are not considered.} For the same reason as that in the 200~percent case, IDMA cannot be effectively operated even in the 300~percent case.

In summary, the SCMA demonstrates the excellent performance over a wide SNR range. Especially, with the aid of a high-complexity MPA receiver, diverse overloading factors and code rates can be applied to the SCMA system. The IDMA shows the best performance with 150 and 200~percents overloading and low code rate. Since the amount of tolerable interference is relatively low, the IDMA works poorly, in contrast, with high overloading factors. Finally, even though the diversity gain of the MUSA is relatively low compared to the gain in other schemes, it provides good BLER performance with low and moderate overloading factors. To support high overloading, the MUSA requires low code rate schemes.

\begin{itemize}
\item {Average Sum Spectral Efficiency}
\end{itemize}

We design modulation and coding scheme (MCS) levels for system-level evaluation, and calculate achievable average sum spectral efficiency according to the MCS levels. The design criterion is to find overloading factors and code rates that satisfy the target BLER=0.1 and maximize the sum user throughput at a given SNR. In the design of MCS levels, we consider a finite set of candidate overloading factors and code rates as follows: \{150, 200, 300~percents\} and $\{0.1,\,0.2,\, \cdots 0.9\}$. Through the Monte-Carlo simulation, we select MCS to maximize spectral efficiency as represented in Figs.~2d-2f. As noted in Figs.~2a-2c, SCMA can operate in a wide-range of overloading factors and code rates, making it possible to achieve an average sum spectral efficiency of up to 4.8~bps/Hz. On the other hand, MUSA shows superior performance when 200~percent users are overloaded up to 2.1~bps/Hz spectral efficiency. The MCS levels for IDMA are composed of a 150~percent overloading factor with code rates 0.1$\sim$0.7, the spectral efficiency of which value is up to 2~bps/Hz.  Even though IDMA and MUSA cannot obtain higher spectral efficiency by raising the overloading factor, they can provide higher throughput than SCMA at the low SNR regime. \vspace{-5mm}

\subsection{System-level Simulation}

\begin{itemize}
\item {System Modeling}
\end{itemize}

Figures~3a and 3b show the simulation environments. We model two realistic 3D digital maps to validate the potential of NR-MA schemes in practice. One is an actual urban area--buildings surrounding Gangnam Station in Seoul, South Korea. The other, for an indoor scenario, is  inside Veritas-C building at Yonsei University, South Korea.
In the urban scenario, we deploy 4~base-stations (BSs) with Katherine antennas typically used in LTE and position 7~sectors in the area of interest. This can be considered a form of a 7-cell hexagonal layout but more practical.
In the indoor scenario, 23~BSs with an omnidirectional antennas are located in the middle of a hall on every floor of the building (about 6 BSs per floor), as illustrated in Fig.~3b.
A massive number of users are uniformly distributed on the ground and on every floor of the building at a density of $10^6/\textrm{km}^2/\textrm{floor}$.

\begin{itemize}
\item {Simulation Procedure}
\end{itemize}

System-level simulations are conducted via the following procedures. First of all, to generate communication links from users to BSs, we utilize a 3D ray-tracing tool, Wireless System Engineering (WiSE), developed by Bell Laboratories. We measure the power-delay-profiles and root-mean-square delays. Based on the measurement, a BS calculates the channel quality of each user. The BS then generates the group sets of users with the same MCS level for a group scheduling.\footnote{Even though the MCS design and scheduling algorithm for the NR-MA are still immature and remain open problems, we can derive meaningful observations from this tractable approach, which considers back-and-forward compatibility. To be specific, the NR-MA systems perform a group scheduling for 8 RBs while the OMA system conduct a user scheduling per one RB. For the OMA system evaluation, we utilize the MCS defined in Table 7.2.3-1, 3GPP TS36.213~\cite{MCS}.} The BS randomly schedules a group for NR-MA, and link-level simulations are performed for the scheduled group. Lastly, we obtain the sum rate per transmission-time-interval (TTI, 1 millisecond).

\begin{figure*}
\centering
\includegraphics[width=\textwidth]{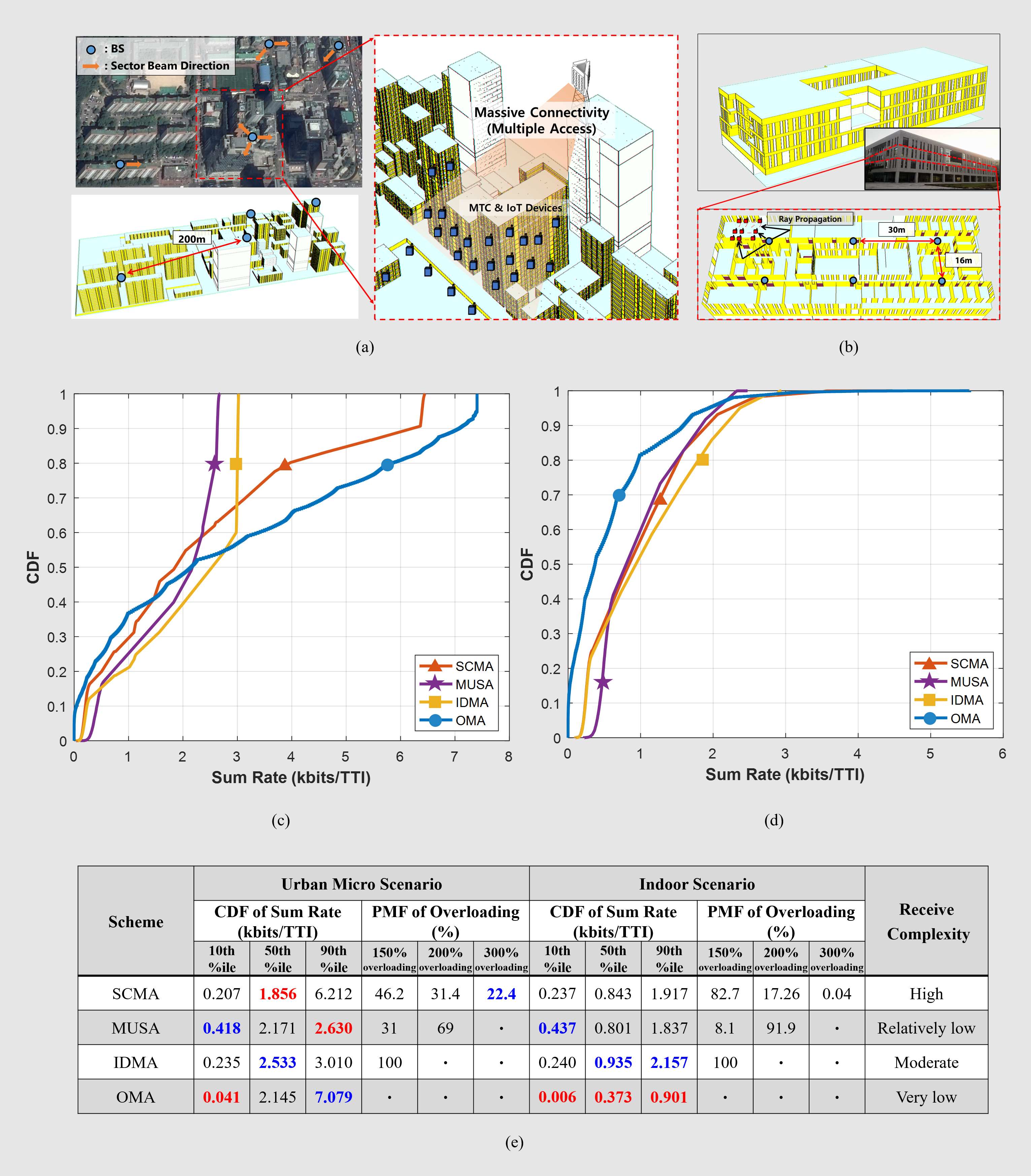}
\caption*{\textbf{Figure 3.} Modeling of practical environments and system-level simulation results: a)-b) the 3D modeling of the Gangnam Station area and Veritas-C building at Yonsei University, respectively; c)-d) CDF of user sum throughput per TTI in urban macro and indoor environments, respectively; e) a summary of performance comparisons.}
\label{fig_3}
\end{figure*}

\begin{itemize}
\item {Simulation Results}
\begin{enumerate}
\item {Full Buffer Traffic Scenario}

\setlength{\parindent}{10pt}Figure~3c shows the system-level performance in the urban micro area. The users' SINRs are widely distributed from -20~dB to 70~dB. In particular, there are lots of strong outdoor-to-outdoor links; as a result, more than 35~percent of users have high SINR ($>$20~dB). In this environment, the NR-MA systems outperform the OMA system at the low and middle SINR regime. The MUSA system with 200~percent significantly improves the sum rate at the low SINR regime. The IDMA achieves superior performance with 150~percent and high code rates at a middle SINR regime. Ultimately, by 300~percent overloading, the SCMA can provide the higher maximum sum rate rather than the other NR-MA schemes. At high SINR regime, however, the OMA system outperforms the NR-MA by using high order modulation techniques. This is because the achievable spectral efficiency of the NR-MA is significantly restricted by the controllable interference amount even at the high SINR regime. The OMA system, on the other hand, can obtain a high peak rate by employing 64QAM and high-rate coding schemes.

\setlength{\parindent}{10pt}The inside of a building, meanwhile, is a contrasting type of environment. We assume that there are 23 dense sensor networks in the building with the width of 60~m, length of 15~m, and height of 16~m. Due to the interference-limited environment, devices suffer from strong interferences caused by neighboring massive devices. Consequently, almost all users have low- and middle-range SINRs (-10$\sim$15~dB). Figure~3d depicts the cumulative distribution function (CDF) of the user sum rate in the indoor scenario. As shown in the figure, the NR-MA schemes significantly enhance the sum capacity. Also, the NR-MA schemes are able to achieve a peak sum rate nearly equivalent to that of the OMA. In terms of comparing the candidates, the overall tendency is similar to the urban micro scenario: MUSA in the 10th percentile and IDMA in the 50th and 90th percentiles show superior performance, respectively. A summary of these evaluation results can be found in~Fig.~3e.

\begin{figure}[!t]
\centering
\includegraphics[width=\columnwidth]{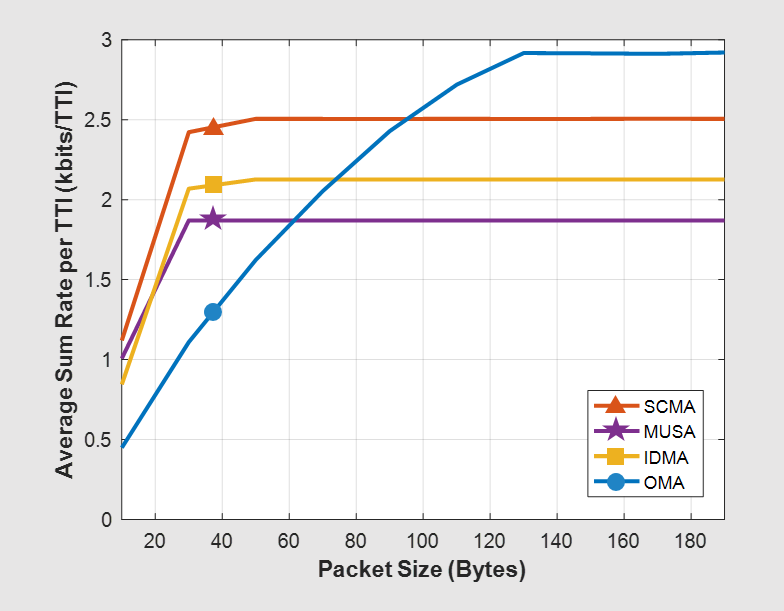}
\caption*{\textbf{Figure 4.} Average sum rate performance with various packet sizes in the urban micro area.}
\label{fig_4}
\end{figure}

\item {Non-full Buffer Traffic Scenario}

\setlength{\parindent}{10pt}The number of target applications of the 5G system has been rapidly growing, and data transmission patterns have become much more diverse. In this regard, the 3GPP standardization groups determined target packet sizes from 20~bytes to 200~bytes for the mMTC scenario. Figure~4 presents the average sum rate performance of the NR-MA systems. With the smaller packet sizes, the NR-MA systems obtain the greater sum rate gains than does the OMA system. When the packet is small enough to be contained in one resource block (RB), there are unused resource elements (REs) in the RB, which lead to inefficient resource utilization. In this case, the NR-MA systems significantly improve spectral efficiency by overloading a large number of users' packets into 8~RBs. From a time resource perspective, it also means that the NR-MA systems can deal with the given traffic faster than the OMA. With high overloading, SCMA in particular is the most favorable for burst small packet transmissions in the massive uplink scenario.
\end{enumerate}

\end{itemize}

\subsection{Insights for System Design}

From the link-level and system-level evaluations, we observe that the candidate schemes of NR-MA tremendously improved the sum rate performance in massive uplink communication systems. In particular, the NR-MA systems are specialized for small packet transmissions in interference-limited environments such as crowded sensor networks. Moreover, in the NR-MA systems, signaling overhead can be reduced by replacing user-specific information with group-specific information. For example, the NR-MA systems can perform group scheduling which enables users within a group to utilize the same transmission parameters such as the positions of scheduled RBs, the MCS level, and the number of repetitions. To support this group-basis operation, a group cell radio network temporary identifier (group c-RNTI) for NR-MA should be newly investigated in 5G networks. In the initial random access stage, instead of c-RNTI, a BS can assign a group c-RNTI based on the users' channel qualities and controllable interference. Through group physical downlink control channel (PDCCH) with a group c-RNTI, common downlink control information (DCI) can be delivered to the group. In addition, since the aggregation level of DCI can be determined up to the spreading factor of NR-MA, the aggregation of DCI allows cell edge users to decode DCI more reliably.

Another noteworthy point is that no single NR-MA scheme overwhelms the others including the OMA in various environments. Fortunately, a massive number of users coexist in 5G networks for various applications with different requirements. Therefore, a smart multiple access strategy is needed that opportunistically operates according to target scenarios. In a situation that requires high data rate communications with a small number of users, such as urban cellular networks, the OMA schemes can be adopted. For applications that support a large number of concurrent small packet transmissions like smart metering networks, the codebook-based MA with a powerful MPA can be beneficial to highly overload data. The sequence-based MA can be more advantageous for areas that require many link connections in addition to coverage enhancement. Also, the interleaver/scrambler-based MA can be applied to an interference-limited area with a heavy traffic load. Through flexible system operation, we anticipate NR-MA schemes to be a promising capacity booster for 5G.

\section{Challenges}

While we have discovered that system capability is greatly enhanced by NR-MA schemes, we recognize several research challenges remaining to be resolved.

\emph{\textbf{Resource and MA-signature Allocation}}: Since we target NR-MA systems to deal with massive link connections, handling dynamic user scheduling is difficult. More efficient resource management techniques are needed. We can consider two alternatives: i) grant-free access and ii) group scheduling-based access. The grant-free access in particular needs no explicit scheduling permission from the BSs. It does need, however, an advanced collision recovery and a user behavior detection algorithm. In the group scheduling case, group management protocol should be investigated with consideration of transmission patterns and target quality-of-service (QoS). Another critical issue is the MA-signature allocation. A pre-configured MA-signature allocation or random MA-signature selection can be regarded as the candidate approach. Also, to implement the NR-MA systems, researchers should study the handling of MA-signature collisions.

\emph{\textbf{Link Adaptation}}: Most prior work has focused on scenarios in which all the overloaded users transmit with the same MCS. This simple group-basis operation has advantages in terms of overhead reduction. The strong restriction of MCS, however, might be an obstacle to achieving the theoretical sum capacity of an uplink multi-user channel. Furthermore, the BSs have to discover user groups with the same channel quality, thus reducing the flexibility of system operation. To resolve these problems, before providing guidelines for the NR-MA system operation, researchers should first theoretically analyze user overloading and achievable rates. In addition, since the lengths and components of MA-signatures significantly impact diversity gain and the interference averaging effect, researchers may consider a new concept of link adaptation to be signature re-assigning algorithms. Based on comprehensive analyses, researchers should investigate a specialized link adaptation for NR-MA systems.

\emph{\textbf{Channel Estimation}}: In the MUD receivers of NR-MA schemes, overall decoding performance relies heavily on the accuracy of channel information. For a reliable channel estimation, as many pilots as the number of overloaded users are needed. There is a heavy burden to allocating resources orthogonally for each user's pilot in the mMTC scenario. For the resource minimization effort of overheads, researchers should study quasi-orthogonal or non-orthogonal pilots with advanced channel estimation techniques such as MMSE with interference cancellation.

\emph{\textbf{Synchronization}}: According to the agreement made during the 3GPP RAN1 \#85 meeting, the uplink synchronization for the grant-free NR-MA is assumed to be the same as the downlink transmission timing. In realistic environments, however, this assumption could not always be guaranteed due to the users' geometrical position differences. Without a timing advance process, there may arise situations in which timing offsets between users are greater than the length of a cyclic prefix. Since the synchronization problem is also directly related to the MUD complexity as well as to the decoding performance, researchers should investigate receiving techniques and frame structures that are robust to timing errors.

\section{Conclusion}
This article has investigated multiple access technologies for 5G new radio. We have discussed the basic principles, categorization, and key features of transceiver structures. Through the link-level simulation, it has been observed that the tolerable amount of inter-user interference could be affected by the characteristics of MA-signatures and receiving algorithms. Further, by modeling the realistic urban and indoor areas with a 3D ray tracing tool, we have verified the great potential and feasibility of new multiple access schemes. From these evaluations, we have discovered superior conditions for each scheme, and drawn insights for multiple access system operation. Lastly, we have introduced key challenges for implementing NR-MA, such as resource management, link adaptation, channel estimation, and synchronization. We expect our investigation will assist for the NR-MA schemes in gaining a promising technical position for next generation wireless networks.



\begin{IEEEbiography}{Hyunsoo Kim} (S'12)
received the B.S. degrees from the School of Electrical and Electronic Engineering, Yonsei university, Seoul, Korea, in 2012. He also received the Scholarship of National Research Foundation of Korea during his B.S studies. He is working toward the Ph.D. degree in Electrical Engineering at Yonsei University. His research interests are new waveform, non-orthogonal multiple access, licensed assisted access, and full-duplex.
\end{IEEEbiography}

\begin{IEEEbiography}{Yeon-Geun Lim} (S'12)
received his B.S. degree in Information and Communications Engineering from Sungkyunkwan University, Korea in 2011. He is now
with the School of Integrated Technology, Yonsei University, Korea and is working toward the Ph.D. degree. His research interest includes massive MIMO, new waveform, full-duplex, and system level simulation for 5G networks.
\end{IEEEbiography}

\begin{IEEEbiography}{Chan-Byoung Chae}(SM'12)
is Underwood Distinguished Professor in the School of Integrated Technology, Yonsei University, Korea. Before joining Yonsei University, he was with Bell Labs, Alcatel-Lucent, Murray Hill, New Jersey, as a Member of Technical Staff, and Harvard University, Cambridge, Massachusetts, as a Post-doctoral Research Fellow. He received his Ph.D. degree in Electrical and Computer Engineering from The University of Texas at Austin, TX, USA in 2008.

He was the recipient/co-recipient of the Yonam Research Award from LG Foundation (2016), the Best Young Professor Award from Yonsei University (2015), the IEEE INFOCOM Best Demo Award (2015), the IEIE/IEEE Joint Award for Young IT Engineer of the Year (2014), the KICS Haedong Young Scholar Award (2013), the IEEE Signal Processing Magazine Best Paper Award (2013), the IEEE ComSoc AP Outstanding Young Researcher Award (2012), the IEEE Dan. E. Noble Fellowship Award (2008), two Gold Prizes (1st) in the 14th/19th Humantech Paper Contest. He currently serves as an Editor for the \emph{IEEE Comm. Mag.}, the \emph{IEEE Trans. on Wireless Comm.}, the \emph{IEEE Wireless Comm. Letters}, the \emph{IEEE/KICS Jour. Comm. Nets}, and the \emph{IEEE Trans. on Molecular, Biological, and Multi-scale Comm.}
\end{IEEEbiography}

\begin{IEEEbiography}{Daesik Hong} (S'86, M'90, SM'05)
received the B.S. and M.S. degrees in Electronics Engineering from Yonsei University, Seoul, Korea, in 1983 and 1985, respectively, and the Ph.D. degree from the School of Electronics Engineering, Purdue University, West Lafayette, IN, in 1990. He joined Yonsei University in 1991, where he is currently the Dean of the College of Engineering and a Professor with the School of Electrical and Electronic Engineering. He is also currently President of the Institute of Electronics and Information Engineering, Korea. He has been serving as Chair of Samsung-Yonsei Research Center for Mobile Intelligent Terminals. He also served as a Vice-President of Research Affairs and a President of Industry-Academic Cooperation Foundation, Yonsei University, from 2010 to 2011. He also served as a Chief Executive Officer (CEO) for Yonsei Technology Holding Company in 2011, and served as a Vice-Chair of the Institute of Electronics Engineers of Korea (IEEK) in 2012.

Dr. Hong is a senior member of the IEEE. He served as an editor of the IEEE Transactions on Wireless Communications from 2006 to 2011. He currently serves as an editor of the \emph{IEEE Wireless Comm. Letters}. He was appointed as the Underwood/Avison distinguished professor at Yonsei University in 2010, and received the Best Teacher Award at Yonsei University in 2006 and 2010. He was also a recipient of the Hae-Dong Outstanding Research Awards of the Korean Institute of Communications and Information Sciences (KICS) in 2006 and the Institute of Electronics Engineers of Korea (IEEK) in 2009. His current research activities are focused on future wireless communication including new waveform, non-orthogonal multiple access, full-duplex, energy harvesting, and vehicle-to-everything communication systems. More information about his research is available at http://mirinae.yonsei.ac.kr.
\end{IEEEbiography}

\end{document}